\documentclass[twocolumn,showpacs,preprintnumbers,amsmath,amssymb]{revtex4}
\usepackage{amssymb}
\usepackage{graphicx,color}% Include figure files
\usepackage{bm}% bold math

\begin{document}

\title{Evolution Effects on Parton Energy Loss with Detailed Balance}

\author{Luan Cheng$^{a,b,c}$ and Enke Wang$^{a,b}$}

\affiliation{$^a$Institute of Particle Physics, Huazhong Normal
University, Wuhan 430079, China\\
$^b$Key Laboratory of Quark $\&$ Lepton Physics (Huzhong Normal
University), Ministry of Education, China\\
$^c$Institut f\"ur Theoretische Physik, Goethe Universit\"at,
Max-von-Laue Str.\ 1, D-60438, Frankfurt am Main, Germany}

%\date{\today}

\begin{abstract}
The initial conditions in the chemical non-equilibrated medium and
Bjorken expanding medium at RHIC are determined. With a set of rate
equations describing the chemical equilibration of quarks and gluons
based on perturbative QCD, we investigate the consequence for parton
evolution at RHIC. With considering parton evolution, it is shown
that the Debye screening mass and the inverse mean free-path of
gluons reduce with increasing proper time in the QGP medium. The
parton evolution affects the parton energy loss with detailed
balance, both parton energy loss from stimulated emission in the
chemical non-equilibrated expanding medium and in Bjorken expanding
medium are linear dependent on the propagating distance rather than
square dependent in the static medium. The energy absorption can not
be neglected at intermediate jet energies and small propagating
distance of the energetic parton in contrast with that it is
important only at intermediate jet energy in the static medium. This
will increase the energy and propagating distance dependence of the
parton energy loss and will affect the shape of suppression of
moderately high $P_{T}$ hadron spectra.
\end{abstract}


\maketitle

\section{Introduction}
One of the challenging goals of heavy-ion physics is to detect
quark-gluon plasma (QGP). The two nuclei pass through each other,
interact, and then produce a dense plasma of quarks and gluons. As
the initial parton density is large and the partons suffer many
collisions in a very short time, the initial partonic system may
attain kinetic equilibrium. But does it attain chemical equilibrium?
This question has been investigated in the framework of parton
cascade model \cite{Geiger:1995sf}, which is based on the concept of
inside-outside cascade
\cite{Anishetty:1980sf,Hwa:1986sf,Balizt:1987} and evolve parton
distributions by Monte-Carlo simulation of a relativistic transport
equation involving lowest order perturbative QCD scattering and
parton fragmentations. From the numerical studies
\cite{Geiger:1993sf,Biro:1994sf,Geiger:1992sf,Kawrakow:1992sf,Shuryak:1992sf,Muller:1992sf}
three distinct phases of parton evolution can be distinguished: (1)
Gluon thermalize very rapidly, reach approximately isotropic
momentum space distribution after a time of the order of 0.3 fm/c.
(2) Full equilibration of gluon phase space density takes
considerably longer. (3) The evolution of quark distributions lags
behind that of the gluons because the relevant QCD cross sections is
suppressed by a factor of 2-3. This calculation indicates that the
QGP likely to be formed in such collisions are far from chemical
equilibrium.

Gluon radiation induced by multiple scattering for an energetic
parton propagating in a dense medium leads to induced parton energy
loss or jet quenching. Jet quenching is manifested in both the
suppression of single inclusive hadron spectra at high transverse
momentum $p_T$ region \cite{phenix} and the disappearance of the
typical back-to-back jet structure in dihadron correlation as
discovered in high-energy heavy-ion collisions at RHIC \cite{star}.
The theoretical investigation of jet quenching has been widely
carried out in recent years
\cite{GW94,BDPMS,Zakharov,GLV,Wiedua,GuoW}. It is found that in the
static medium the radiative energy loss is proportional to square of
propagating distance. Later the detailed balance effect with gluon
absorption was included. It has been shown that the gluon absorption
play an important role for intermediate jet energy region in the
static medium \cite{Wang:2001sf}. But the plasma is not static, it
will expand, cool and become more dilute. Recently medium expansion
is included in the jet energy loss by stimulated emission in Ref.
\cite{Gyulassy:2002sf}. However, only the gluon distribution
evolution in a thermodynamical equilibrated expanding medium is
considered, the temperature evolution, which play an important role
for debye screened mass calculation and affects the mean free path
and opacity, was neglected in Ref. \cite{Gyulassy:2002sf}. Moreover,
since the QGP is like to be far from chemical equilibrium, the
effect of parton chemical equilibration on jet energy loss need to
be studied.

In this paper, we will study the effects of temperature and fugacity
(chemical potential) evolution on jet quenching with detailed
balance. In general, the question of thermodynamical equilibration
can be decided with microscopic transport models [5]. In this paper,
we study a simpler problem which allows us to use a macroscopic
model to get the initial conditions and parton evolution. Our
strategy will be to assume that the parton distribution can be
approximately by thermal phase space distribution with
non-equilibrium fugacity $\lambda_g$ and $\lambda_q$ of gluon and
quark, use a set of rate equations to describe the chemical
equilibration of partons, compare $dE_T/dy$ and $dN/dy$ which we get
with the data from RHIC, we can then determine the initial
conditions of proper time, fugacity, and the time dependence of the
parameters $T$, $\lambda_g$ and $\lambda_q$. We will also study the
a thermal-equilibrated and chemical-equilibrated medium expansion -
Bjorken expansion with $\lambda_{g(q)}=1$ and compare the
difference. With these results, we obtain the Debye screening mass,
mean free path, and opacity from the perturbative QCD at finite
temperature in a thermal equilibrated, but chemical non-equilibrated
medium and find that they are different with those in
thermodynamical medium and static medium. Then, we will investigate
the evolution effects on both the final-state radiation associated
with the hard processes and the radiation induced by final-state
multiple scattering in the medium.

\section{Parton Equilibration at RHIC}\label{sec:a}

\subsection{Basic Equations}
We consider here a thermal equilibrated, but chemical
non-equilibrated system, and assume that the parton distribution can
be approximated by thermal phase space distributions with
non-equilibrium fugacities $\lambda_i$ which gives the measure of
derivation from chemical equilibrium,
\begin{equation}
f(k;T,\lambda_i) = \lambda_i(e^{\beta u\cdot
k}\pm\lambda_i)^{-1}\label{one}\, ,
\end{equation}
where $\beta$ is the inverse temperature and $u$ is the
four-velocity of the local moving reference frame, $i=g, q, {\bar
q}$ for gluon, quark, anti-quark, respectively.

As discussed in Ref. \cite{Biro:1994sf}, as an approximation the
momentum distributions can be written in the factorized Bose or
Fermi-Dirac form
\begin{equation}
f(k;T,\lambda_i) = \lambda_i(e^{\beta u\cdot k}\pm
1)^{-1}\label{two}\, ,
\end{equation}
which we will adopt in most of the following calculation.

In general, chemical reactions among partons can be quite
complicated because of the possibility of initial and final-state
gluon radiations. However, since interference effects due to
multiple scatterings inside a dense medium lead to a strong
suppression of soft gluon radiation, we shall only consider
processes in which a single addition gluon is radiated, such as
$gg\rightarrow ggg$. But in order to permit approach to chemical
equilibrium, we should also consider the reverse process. Closer
inspection shows that gluon radiation is dominated by the process
$gg\rightarrow ggg$, because radiative processes involving quarks
have substantially smaller cross sections in pQCD, and quarks are
considerably less abundant than gluons in the initial phase of the
chemical evolution of the parton gas. Here we are interested in
understanding the basic mechanisms, so we restrict our consideration
to the dominant reaction mechanisms for the equilibration of each
parton flavor. These are four processes \cite{Biro:1994sf}:
\begin{equation}
gg\leftrightarrow ggg,\qquad gg\leftrightarrow q\bar{q}\, .
\end{equation}

Restricting to reactions, the evolution of the parton densities is
governed by the master equations,
\begin{eqnarray}
\partial_\mu(\rho_{g}u^{\mu})&=&\rho_{g}R_3(1-\lambda_{g})
-2\rho_{g}R_2(1-\frac{\lambda_{q}\lambda_{\bar{q}}}
{\lambda_{g}^2})\, ,
\\
\partial_\mu(\rho_{q}u^{\mu})&=& \rho_{q}R_2(1-\frac{\lambda_{q}
\lambda_{\bar{q}}}{\lambda_{g}^2})\, ,
\end{eqnarray}
where $R_2=\frac{1}{2}\sigma_{2}n_g$,
$R_3=\frac{1}{2}\sigma_{3}n_g$, $\sigma_2$ and $\sigma_3$ are
thermally averaged velocity weighted cross sections,
$\sigma_2=\langle\sigma(gg\rightarrow q\bar{q})v\rangle$ and
$\sigma_3=\langle\sigma(gg\rightarrow ggg)v\rangle$. It has been
calculated that
$R_2\approx0.064N_f\alpha_{s}^2\lambda_{g}T\ln(7.5/\lambda_{g})^2$
and $R_3=2.1\alpha_{s}^{2}T(2\lambda_g-\lambda_g^2)^{1/2}$ in Refs.
\cite{Biro:1994sf,Wang:1997sf}.

Using Bose and Fermi-Dirac momentum distributions, the gluon and
quark densities are related to temperature as

\begin{eqnarray}
\rho_{g}&=& \frac{16}{\pi^2}\zeta(3)\lambda_{g} T^3 ,
\\
\rho_{q}&=& \frac{9}{2\pi^2}\zeta(3)N_{f} \lambda_{q} T^3,
\end{eqnarray}
where $\zeta(3)\approx 1.2$, $N_f$ is the number of dynamical quark
flavors.

If we assume that parton scatterings are sufficiently rapid to
maintain local thermal equilibrium, and therefore we can neglect
effects of viscosity due to elastic and inelastic scatterings, we
can have the hydrodynamic equation,
\begin{equation}
\partial_{\mu}(\varepsilon u^{\mu})+P\partial_{\mu}u^{\mu}=0\, ,
\end{equation}
where $\varepsilon$ and $P$ are energy density and pressure of the
hot medium.

In order to obtain analytical solutions, we will neglect the
transverse expansion and consider only a purely longitudinal
expansion of the parton plasma, so Eq.(8) can be rewritten as
\begin{equation}
\frac{d\varepsilon}{d\tau}+\frac{\varepsilon+P}{\tau}=0\, .
\end{equation}
With the additional constraint of the baryon number conservation, we
get
\begin{equation}
\partial_{\mu}(\rho_{_B} u^{\mu})=0\, .
\end{equation}

Once the initial conditions are obtained, the evolution of
temperature $T(\tau)$, and fugacity $\lambda_g(\tau)$,
$\lambda_q(\tau)$ can be determined by solving the rate Eqs.(4),
(5), (9) and (10) together. So the input of the initial condition
plays an important role to investigate the effects of the evolution
for the parton system.

\subsection{Initial Conditions}

\subsubsection{Transverse Energy}

In order to solve the rate equations discussed in the last section,
one has to specify the initial conditions.

The transverse energy per unit rapidity can be expressed as
\cite{Xu:2005sf}
\begin{equation}
\frac{dE_T}{dy}=\tau A\int d\eta d^{2}p_{T}p_{T}^2cosh\xi
f(k;T,\lambda_i)\, .
\end{equation}
where $\xi=y-\eta$, $y$ and $\eta$ are the rapidity and
pseudorapidity, respectively.

Using the momentum distribution discussed in last section, the
transverse energy per unit rapidity can be deduced as
\begin{eqnarray}
\frac{dE_T}{dy}&=&[g_{f}(\lambda_{q}+\lambda_{\overline{q}})+g_{b}\lambda_g]\frac{\tau\pi
R^2 }{2\pi^3}\int d\eta d^{2}p_{T}p_{T}^2 \nonumber \\
&& \times cosh\xi e^{-\frac{p_{T}cosh\xi}{T(\tau)}}\, .
\end{eqnarray}

\begin{figure}
\includegraphics[width=8cm]{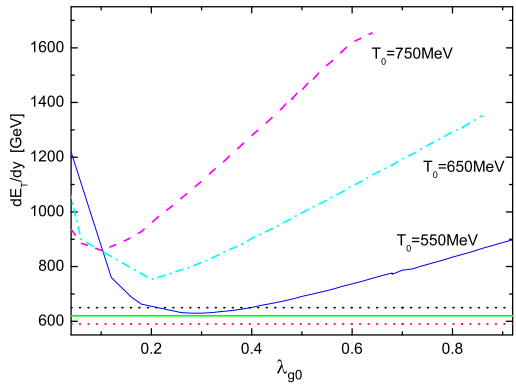}
\caption[a]{ The initial fugacity $\lambda_{g0}$ dependence of
transverse energy per unit rapidity for the most central events. The
solid, dash-dot and dash curves show the results of $dE_T/dy$ for
initial proper time $\tau_0=0.3$ $fm$ and initial temperature
$T_0=550$ $MeV$, 650 $MeV$, 750 $MeV$ at the freezing out time,
respectively. The area between the two dot line is the data
$dE_T/dy|_{y=0}=620\pm33$ $GeV$ for the 5\% most central events of
Au-Au collisions from RHIC\cite{Adams:2004sf}.} \label{fig:detdy1}
\end{figure}

For the anti-quark fugacity,
$\lambda_{\overline{i}}=\lambda_i^{-1}$. We take $\tau_0=0.3$ $fm$
as in Ref. \cite{Biro:1994sf}. The QGP medium freezes out when the
temperature reduces to about 160-170 $MeV$. With different initial
temperature $T_0$, we will get different parton equilibration and
then different curves of transverse energy per unit rapidity
$dE_{T}/dy$ as the function of $\lambda_{g0}$ at the freezing out
time (4.4 $fm$) as shown in Fig. \ref{fig:detdy1}.

The data from RHIC shows that at $\sqrt{s}=200A$ $GeV$ transverse
energy per unit rapidity is $dE_T/dy|_{y=0}=620\pm33$
$GeV$\cite{Adams:2004sf} for the 5\% most central events of Au-Au
collisions. It is found that curve of the transverse energy per unit
rapidity doesn't intersect the straight line $dE_T/dy|_{y=0}=620$ $
GeV$ if the initial temperature is higher than $550$ $MeV$, only
when the initial temperature is $T_0=550$ $MeV$, $\lambda_{g0}=0.3$,
the curve is tangential to the straight line.
$dE_T/dy|_{y=0}=620\pm33$ $GeV$\cite{Adams:2004sf} for the 5\% most
central events of Au-Au collisions is measured after the time freeze
out. After freezing out hard scattering seldom happens so that there
is little transverse momentum transfer and $dE_T/dy|_{y=0}$ is
stable. Here only tangential point is the minimum of the curve and
stable. If the initial temperature is lower than $550$ $MeV$ they
have two points of intersection, which, however, are not stable. It
implies that the initial temperature is $550$ $MeV$, $\lambda_{g0}$
is $0.3$ for the central events of Au-Au collisions at RHIC. Using
the same method, we obtain that the initial temperature is $420$
$MeV$ for Bjorken expansion($T^3\tau=T_0^3\tau_0$) for thermal and
chemical equilibrium system.

\subsubsection{Particle Multiplicities}

Secondly we demonstrate that the initial condition determined above
is consistent with that from the particle multiplicities.

The thermodynamic functions for a many-particle system for an
ensemble at temperature $T$ and fugacity $\lambda$ can be derived
from the grand partition,
\begin{eqnarray}
lnZ_F&=&\frac{g_{F}V}{6\pi^2T}\int_{0}^{\infty}dp\frac{p^4}{(p^2-m^2)^{1/2}}
[\frac{1}{1+\lambda^{-1}e^{\beta\sqrt{p^2+m^2}}} \nonumber
\\
&&+\frac{1}{1+\lambda e^{\beta\sqrt{p^2+m^2}}}]
\label{lnzf}
\end{eqnarray}
for fermions and
\begin{equation}
lnZ_B=\frac{g_{B}V}{6\pi^2T}\int_{0}^{\infty}dp\frac{p^4}{(p^2-m^2)^{1/2}}
\frac{1}{\lambda^{-1}e^{\beta\sqrt{p^2+m^2}}-1}
\label{lnzb}
\end{equation}
for bosons, where $g_{_F}$ and $g_{_B}$
are the degeneracy factors of fermions and bosons, $V$ is the volume
of the parton gas.

If we neglect the mass for light quarks and gluons, the momentum
integral Eq. (\ref{lnzf}), (\ref{lnzb}) can be calculated exactly.
For light quarks and gluon the grand partition functions can be
expressed as
\begin{eqnarray}
TlnZ_{q(\bar{q})} &= &\frac{g_q
V}{\pi^2}(\lambda_q+\lambda_{q}^{-1})T^4\, ,
\label{tlnzq}\\
TlnZ_{g} &=& \frac{g_g V}{\pi^2}\lambda_{g}T^4\, .
\label{tlnzg}
\end{eqnarray}

In order to get the expressions for the entropy density and particle
number density, we recall the first law of thermodynamics
\begin{equation}
E=F(V,T,\mu)+TS(V,T,\mu)+\mu N((V,T,\mu)\, ,
\label{ee}
\end{equation}
where $F, T, S, \mu, N$ are the free energy, temperature, entropy,
chemical potential and particle number, respectively.

We can evaluate the free energy and the average value of particle
number for the grand-canonical partition function. With the
thermodynamics Eq.(\ref{ee}), we obtain the entropy density in the
chemical non-equilibrated as
\begin{eqnarray}
s &=&
\frac{48}{\pi^2}(\lambda_q+\lambda_{q}^{-1})T^3+\frac{64}{\pi^2}\lambda_{g}T^3
   \nonumber\\
   &&-ln\lambda_q(\lambda_q-\lambda_{q}^{-1})\frac{12}{\pi^2}T^3
   -(ln\lambda_g)\lambda_g\frac{16}{\pi^2}T^3\, .
\end{eqnarray}

In hydrodynamics, the relation between entropy $S$ and particle
number $N$ is $S/N= 4$. In a thermal equilibrated but chemical
non-equilibrated system here, with considering the thermodynamics
Eq.(\ref{ee}) and the grand canonical partition functions Eqs.
(\ref{tlnzq}) and (\ref{tlnzg}), we can deduce
\begin{equation}
\frac{S}{N}=\frac{E+PV-\mu N}{TN}=\frac{s_i}{dN_i/dy/\tau \pi
R^2}=4-ln\lambda_i\, .
\end{equation}
where $s_i$ and $dN_i/dy$ are the entropy density and particle
number distribution for quarks, anti-quarks, and gluons.

So we have
\begin{eqnarray}
dN/dy &=& \sum_i dN_i/dy  \nonumber \\ &=& \frac{\tau\pi
R^2}{4-ln(\lambda_q)}[\frac{48}{\pi^2}
   \lambda_q T^3-(ln\lambda_q)\lambda_q\frac{12}{\pi^2}T^3]\nonumber
   \\
   &&+\frac{\tau\pi R^2}{4-ln(\lambda_q^{-1})}[\frac{48}{\pi^2}
   \lambda_q^{-1}
   T^3-(ln\lambda_q^{-1})\lambda_q^{-1}\frac{12}{\pi^2}T^3]\nonumber
   \\
   &&+\frac{\tau\pi R^2}{4-ln(\lambda_g)}[\frac{64}{\pi^2}
   \lambda_g T^3-(ln\lambda_g)\lambda_g\frac{16}{\pi^2}T^3],
\end{eqnarray}
where $R$ is the transverse radius of the medium and taken as $6.5$
$fm$, which is approximately the size of radius of Au.

\begin{figure}
\includegraphics[width=8cm]{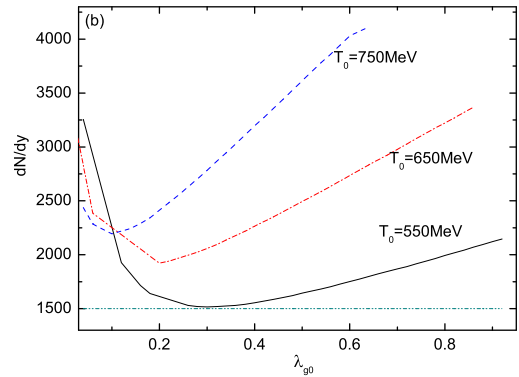}
\caption[a]{The initial fugacity $\lambda_{g0}$ dependence of
particle multiplicities. The solid, dash-dot and dash curves show
the results for initial proper time $\tau_0=0.3$ $fm$ and initial
temperature $T_0=550$ $MeV$, 650 $MeV$, 750 $MeV$, respectively. The
dot line indicate $dN/dy=1500$ from Ref. \cite{Gyulassy:2002sf}.}
\label{fig:dndy}
\end{figure}

As discussed in the above subsection, with different initial
conditions $T_0$, $\lambda_{g0}$ and $\lambda_{q0}$, we obtain the
different particle multiplicities $dN/dy$ as shown in Fig.
\ref{fig:dndy}. The data from RHIC shows that particle
multiplicities $dN/dy=1500$  at $\sqrt{s}=200A$ $GeV$ from Ref.
\cite{Gyulassy:2002sf}. Comparing our calculations with the data
from RHIC, we can conclude that the initial conditions which we get
here are consistent with that from transverse energy.

\subsection{Parton Equilibration}
\begin{figure}
\includegraphics[width=8.2cm]{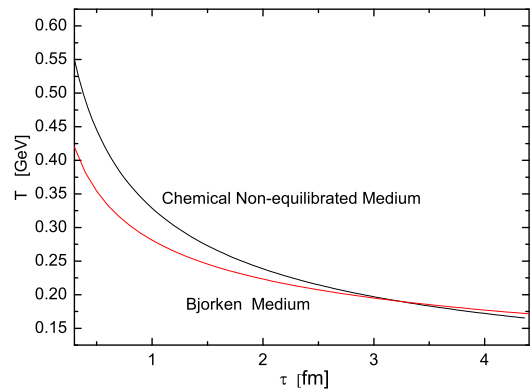}
\caption[a]{Time evolution of the temperature $T$ in chemical
non-equilibrated medium and in Bjorken expanding medium in Au+Au
collisions for 200 GeV/nucleon at RHIC.} \label{fig:Ttau}
\end{figure}

\begin{figure}
\includegraphics[width=7.5cm]{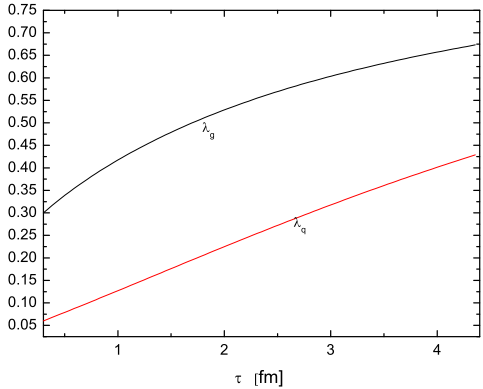}
\caption[a]{Time evolution of the fugacities $\lambda_g$ and
$\lambda_q$ of gluons and quarks in the Au+Au collisions for 200
GeV/nucleon at RHIC.} \label{fig:ltau}
\end{figure}

With the initial conditions obtained above and rate Eqs.(4), (5),
energy conservation Eq.(9) and baryon number conservation Eq.(10),
we can obtain the parton evolution in a chemical non-equilibrated
system. The evolution of temperature and fugacity is shown in Fig.
\ref{fig:Ttau} and Fig. \ref{fig:ltau}. We find that the parton gas
cools faster than predicted in Bjorken expanding medium as shown in
Fig. \ref{fig:Ttau} because the production of additional partons
approaching the chemical equilibrium state consumes an appreciable
amount of energy. Fig. \ref{fig:ltau} shows that fugacity of gluons
$\lambda_g$ and quarks $\lambda_q$ increase with increasing the
proper time $\tau$, until the medium freezes out it hasn't been
totally chemical equilibrated.

\section{Opacity in a Chemical Non-equilibrated QGP Medium}

Parton evolution will lead to the evolution of the Deby screening
mass and parton number density. It will affect the cross section
between the jet and the medium partons, and the opacity when the jet
propagates through the QGP medium.

The Debye screening mass $\mu_D$ is generated by medium effects
\cite{Braaten:1990sf}. Using Bose and Fermi equilibrium
distributions in the above section, one gives the Debye screening
mass,

\begin{eqnarray}
\mu_D^{2} &=& \frac{6g^2}{\pi^2} \int_{0}^{\infty}k f(k) dk \\
&=& \frac{4}{3}\pi\alpha_{s} T^2(\lambda_{q}+2\lambda_{g}).
\end{eqnarray}

\begin{figure}
\includegraphics[width=8.2cm]{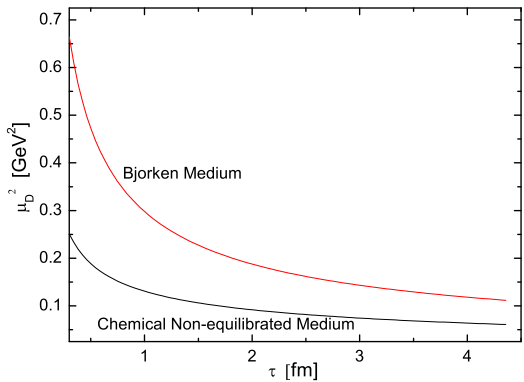}
\caption[a]{Time evolution of the Debye screening mass in chemical
non-equilibrated medium and in Bjorken expanding medium in the Au+Au
collisions for 200 $GeV$/nucleon at RHIC.} \label{fig:mud}
\end{figure}

Using the parton evolution in the above section, we can obtain how
$\mu_D^2$ evolves with the proper time in a chemical
non-equilibrated QGP medium and in Bjorken expanding medium as shown
in Fig. \ref{fig:mud}. We can see that the Debye screening mass in
the chemical non-equilibrated medium is less than that in the
Bjorken expanding medium because of the less fugacity in the
chemical non-equilibrated medium.

The cross sections for the processes, such as $gg\rightarrow gg$ and
$gq\rightarrow gq$, can be calculated from the pQCD, the leading
order elastic scattering cross sections can be expressed as
 \begin{eqnarray} \sigma_{gg} \simeq
\frac{9\pi\alpha_{s}^2}{2\mu_D^2}, \qquad \sigma_{gq} \simeq
\frac{2\pi\alpha_{s}^2}{\mu_D^2}.
 \end{eqnarray}
where $\mu_D$ is the Debye screening mass generated from the medium
 effects with considering both quark and gluon contributions\cite{Wang:2001sf}.

Hence, the mean free-path for a gluon $l_{g}$ is \cite{Wang:1995sf}
\begin{eqnarray}
l_{g}^{-1}&=& \sigma_{gg}\rho_{g}+ \sigma_{gq}\rho_{q} \nonumber\\
&\simeq & \frac{72\alpha_{s}^2}{\mu_D^2\pi}\zeta(3)\lambda_{g}
T^3+\frac{9\alpha_{s}^2}{\mu_D^2\pi}\zeta(3)N_{f} \lambda_{q}T^3.
\end{eqnarray}

\begin{figure}
\includegraphics[width=8.2cm]{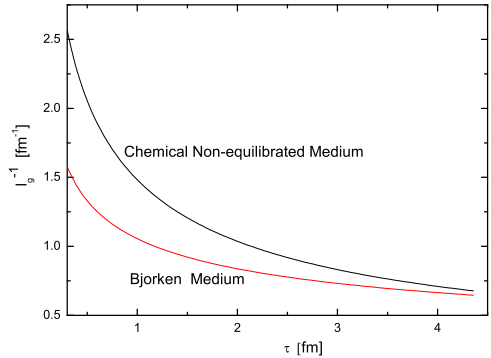}
\caption[a]{Time evolution of the mean free-path for gluon in
chemical non-equilibrated medium and in Bjorken expanding medium in
the Au+Au collisions for 200 $GeV$/nucleon at RHIC.}
 \label{fig:mfp}
\end{figure}

From the above results in the last section of parton equilibration,
one can obtain the proper time dependence of the mean free-path as
shown in Fig. \ref{fig:mfp}, the inverse mean free-path decreases
with increasing proper time, the mean free-path in the chemical
non-equilibrated medium is larger than that in Bjorken expanding
medium due to less Debye screening mass and larger temperature in
the chemical non-equilibrated medium.

We assume that jet travel the distance of $L$ at speed of light in
the chemical non-equilibrated QGP medium, the opacity can be written
as

\begin{equation}
\chi= \int_{\tau_0}^{\tau_0+ L}
(\sigma_{gg}\rho_{g}+\sigma_{gq}\rho_{q})d\tau.
\end{equation}

\begin{figure}
\includegraphics[width=8cm]{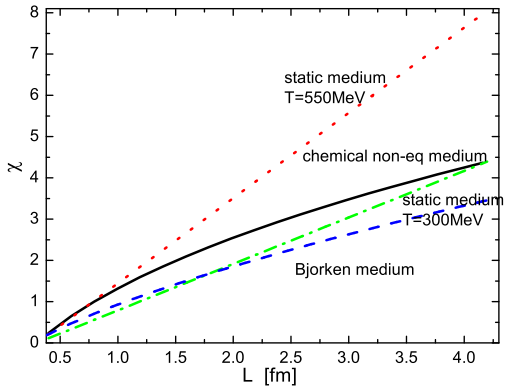}
\caption[a]{Propagating distance dependence of the opacity in the
Au+Au collisions for 200 $GeV$/nucleon at RHIC. The dotted curve is
the opacity in a static medium for temperature $T$=550 $MeV$, and
the dotted dash curve is the opacity in a static medium at
temperature $T$=300 $MeV$. The solid and dash curve indicate that in
the chemical non-equilibrated expanding medium for $T_0$=550 $MeV$
and the Bjorken expanding medium for $T_0$=420 $MeV$, respectively.}
 \label{fig:opa}
\end{figure}

Using the result from parton equilibration in above section, the
propagating distance $L$ dependence of the opacity is shown in
Fig.\ref{fig:opa}. We can see that the opacity increases with
increasing the propagating distance $L$. The opacity at the initial
temperature $T$=550 $MeV$ in the static medium, which means that the
medium always stay at the initial temperature, is larger than that
in the chemical non-equilibrated and Bjorken expanding medium
because of the decrease of the temperature and density in the
expanding medium. In the chemical non-equilibrated expanding QGP
medium, the opacity is larger than that in Bjorken expanding medium
because of the higher temperature and the different ratio of quarks
and gluons in the two mediums. One also finds that the opacity in
the chemical non-equilibrated medium with the initial temperature
$T$=550 $MeV$ is larger than that in a static medium at the
temperature $T$=300 $MeV$ as considered in Ref. \cite{GLV}.

\section{Parton Energy Loss with Detailed Balance}
\label{mode-sec}

The theoretical investigation of jet quenching are mostly carried
out in a static medium and is found that the radiative energy loss
is proportional to the square of propagating distance because of
non-Abelian effect. However, the QGP medium is an expanding medium
with evolution instead of being static. In this section, we will
compare the difference of propagating distance L dependence in the
different medium and the ratio of effective parton energy loss with
and without absorption to study the evolution effects on jet
quenching.

In the leading-log approximation the final-state radiation amplitude
of a quark can be factorized from the hard scattering in an axial
gauge. Taking into account of both stimulated emission and thermal
absorption in the chemical non-equilibrated expanding hot medium,
one obtain the probability of gluon radiation with energy $\omega$
to the 0th order opacity(self quenching and absorption)
\cite{Wang:2001sf}
\begin{eqnarray}
&&\frac{d P^{(0)}}{d \omega}= \frac{\alpha_{s}C_{F}}{2\pi}
\int\frac{dz}{z}\int\frac{dk_{\perp}^2}{k_{\perp}^2}[f_{g}(zE)\delta(\omega+zE)
\nonumber\\
 &&\quad +(1+f_{g}(zE))\delta(\omega-zE)\theta(1-z)]P_{gq}(\frac{\omega}{E}),
\end{eqnarray}
where $C_F$ is the Casimir of the quark jet, $f_{g}(k)$ is the gluon
distribution which has been shown in Eq.(2) and the splitting
function $P_{gq}(z)\equiv P(z)/z=[1+(1-z)^2]/z$ for $q\rightarrow
gq$. The first term corresponds to thermal absorption and the second
term corresponds to gluon emission with the Bose-Einstein
enhancement factor.

Assuming the scale of the hard scattering as $Q^2=4E^2$ and
considering gluon radiation outside a cone with $|k_{\perp}|>\mu_D$,
one has the kinematic limits of the gluon's transverse momentum
\cite{Wang:2001sf},

\begin{equation}
\mu_D^2\leq k_{\perp max}^2\leq 4|\omega|(E-\omega).
\end{equation}

\begin{figure}
\includegraphics[width=8.5cm]{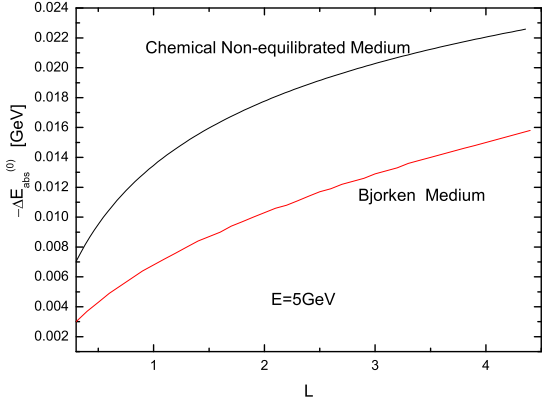}
\caption[a]{Propagating distance dependence of the energy gain
without rescattering in chemical non-equilibrated medium and in
Bjorken expanding medium. } \label{fig:deltaE0}
\end{figure}

Subtracting the gluon radiation spectrum in the vacuum, one then
obtains the energy loss due to final absorption and stimulated
emission when the jet passes through the chemical non-equilibrated
hot medium over the distance $L$,
\begin{eqnarray}
\Delta E_{abs}^{(0)}&=&  \frac{L}{L_0}\int d\omega \omega(\frac{d
P^{(0)}}{d \omega}-\frac{d P^{(0)}}{d
\omega}\mid_{T=0}) \nonumber \\
&=& \frac{\alpha_{s}C_{F}L}{2\pi L_0} \int dz
\int\frac{dk_{\perp}^2}{k_{\perp}^2}[-P(-z)f_{g}(zE) \nonumber
\\ &&+P(z)f_g(zE)\theta(1-z)] \nonumber \\
&=& -\frac{\pi\alpha_s C_F L\langle T
\rangle^2}{3L_0E}[ln\frac{4E\langle T \rangle}{\mu_D^2}
+2-\gamma_E+\frac{\zeta'(2)}{\pi^2}], \nonumber \\
\end{eqnarray}
where $\gamma_{\rm E}\approx 0.5772$, $\zeta^\prime(2)\approx
-0.9376$, $L_0$ is the thickness of the QGP medium, which we
estimate as the diameter of the Au nucleon, and $\langle T \rangle$
is the average temperature during the time the jet propagates
through the medium,
\begin{eqnarray}
<T>=\frac{1}{L} \int _{\tau_0}^{L} T(\tau)d\tau\, .
\end{eqnarray}
As shown in Fig. \ref{fig:deltaE0}, we see that the energy loss
without rescattering increases with increasing $L$ at fixed $E=5$
$GeV$, the energy gain without rescattering in the chemical
non-equilibrated expanding medium is larger than that in the Bjorken
expanding medium because of the higher temperature in the chemical
non-equilibrated medium.

During the propagation of the jet after its production, it will
suffer multiple scattering with targets in the medium. It was shown
by GLV \cite{GLV} that the higher order opacity corrections
contribute little to the radiative energy loss, so we will focus on
the stimulated emission and thermal absorption associated with
rescattering with considering parton equilibration in a chemical
non-equilibrated hot medium. We use the model which is initially
proposed by Gyulassy-Wang(GW) to describe the interaction between
the jet and target partons by a static color-screened Yukawa
potential \cite{Wang:1995sf}. Assuming that a parton is produced at
$\textbf{y}_{0}=(y_{0},\textbf{y}_{0\perp})$ inside the medium with
$y_{0}$ being the longitudinal coordinate, the Yukawa potential is

\begin{eqnarray}
V_{n}&=&2\pi\delta(q^0)v(q_n)e^{-iq_{n}\cdot y_n}
t_{a_n}(j)t_{a_n}(n),
\\
v(q_n)&=&\frac{4\pi \alpha_s}{q_{n}^2+\mu_D^2},
\end{eqnarray}
where $q_n$ is the momentum transfer from a target parton $n$ at the
position $\textbf{y}_{n}=(y_{n},\textbf{y}_{n\perp})$, $t_{a_n}(j)$
and $t_{a_n}(n)$ are the color matrices for the jet and target
parton.

Here we consider the contributions to the first order in the opacity
expansion. By including the interference between the process of the
rescattering and non-rescattering, we obtain the energy loss for
stimulated emission $\Delta E_{rad}^{(1)}$ and energy gain for
thermal absorption $\Delta E_{abs}^{(1)}$ to the first order as
\begin{eqnarray}
\Delta E_{rad}^{(1)} &=& \frac{\alpha_s C_F E}{\pi}
\int_{\tau_0}^{\tau_0+L} d\tau \int dz \int
\frac{dk_{\perp}^2}{k_{\perp}^2} \int d^2 q_{\perp}\nonumber \\
&&\mid \overline{v}(q_{\perp}) \mid^2 \frac{k_{\perp}\cdot
q_{\perp}}{(k_{\perp}-q_{\perp})^2}P(z)(\sigma_{gg}\rho_g+\sigma_{gq}\rho_q)  \nonumber\\
&&<Re(1-e^{i\omega_{1} y_{10}})> \theta(1-z),
\\
\Delta E_{abs}^{(1)} &=& \frac{\alpha_s C_F E}{\pi}
\int_{\tau_0}^{\tau_0+L} d\tau \int dz \int
\frac{dk_{\perp}^2}{k_{\perp}^2} \int d^2 q_{\perp}\nonumber \\
 &&\mid
\overline{v}(q_{\perp}) \mid^2 \frac{k_{\perp}\cdot
q_{\perp}}{(k_{\perp}-q_{\perp})^2}f_{g}(zE)(\sigma_{gg}\rho_g+\sigma_{gq}\rho_q)\nonumber\\
&&\times[-P(-z)) <Re(1-e^{i\omega_{1} y_{10}})>\nonumber \\
&&+P(z)<Re(1-e^{i\omega_{1} y_{10}})>\theta(1-z)].
\end{eqnarray}
where $\omega_{1}=(k_{\perp}-q_{\perp})^2/{2\omega}$, the factor
$(1-e^{i\omega_{1} y_{10}})$ reflects the destructive interference
arising from the non-Abelian LPM effect. Averaging over the
longitudinal target profile is defined as $<\cdots>=\int dy
\rho(y)\cdots$. The target parton number distribution along the jet
direction in the QGP medium is assumed to be an exponential form
$\rho(y)=2exp(-2y/L)/L$. $|{\bar v}({\bf q}_{\perp})|^2$ is the
normalized distribution of momentum transfer from the scattering
centers,
 \begin{eqnarray}
 \label{vbar}
  |{\bar v}({\bf q}_{\perp})|^2 &\equiv& {1\over \sigma_{el}}
  {d^2\sigma_{el}\over d^2{\bf q}_{\perp}}=
  {1\over \pi}{\mu^2_{eff}\over ({\bf q}_{\perp}^2+\mu_D^2)^2}\, ,
 \\
   {1\over \mu^2_{eff}} &=& {1\over \mu_D^2}-{1\over
   q_{\perp max}^2+\mu_D^2}\,\,\, , q_{\perp max}^2\approx 3E\mu_D.
 \end{eqnarray}

\begin{figure}
\includegraphics[width=8.6cm]{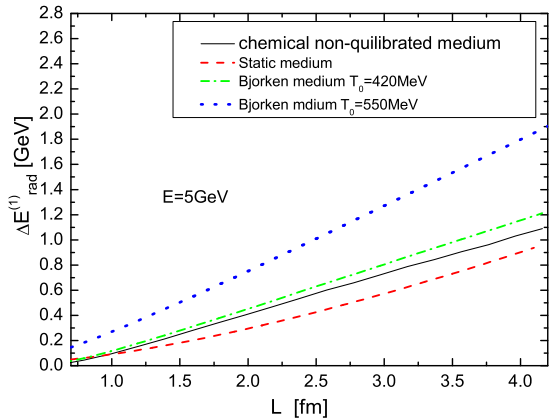}
\caption[a]{Propagating distance dependence of the energy loss
 to the first order
opacity. For static medium case, $T=300$ $MeV$, $\mu_D=0.5$ $GeV$
and $l_{g}=1$ $fm$. For chemical non-equilibrate medium case,
$T_0=550$ $MeV$, $\lambda_{g0}=0.3$ and $\tau_0=0.3$ $fm$. For the
Bjorken expanding case with two different initial temperature,
$T_0=420$ $MeV$ and $T_0=550$ $MeV$, $\tau_0=0.3$ $fm$. }
\label{fig:deradcompare}
\end{figure}

\begin{figure}
\includegraphics[width=8.6cm]{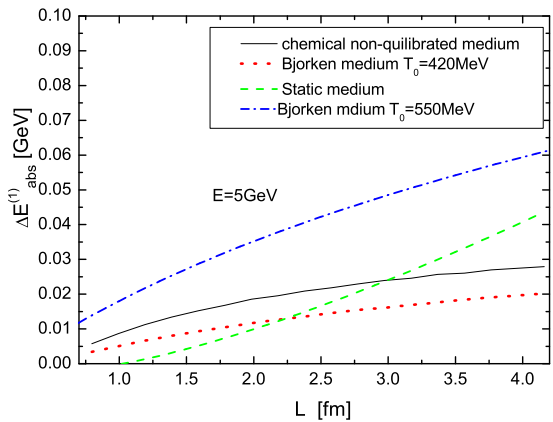}
\caption[a]{Propagating distance dependence of the energy gain to
the first order. For static medium case, $T=300$ $MeV$, $\mu_D=0.5$
$GeV$ and $l_{g}=1$ $fm$. For chemical non-equilibrate medium case,
$T_0=550$ $MeV$, $\lambda_{g0}=0.3$ and $\tau_0=0.3$ $fm$. For the
Bjorken expanding case, $T_0=420$ $MeV$, $\tau_0=0.3$ $fm$.  }
\label{fig:deabscompare}
\end{figure}

The propagating distance dependence of the energy loss and energy
gain for stimulated emission and thermal absorption is shown in Fig.
\ref{fig:deradcompare} and Fig. \ref{fig:deabscompare} at fixed jet
energy $E=5$ $GeV$.

In the limit of $EL>>1$ and $E>>\mu_D$, the approximate asymptotic
behavior of the energy loss of stimulated emission \cite{GLV} and
energy gain of thermal absorption \cite{Wang:2001sf} in a static
medium are

\begin{equation}
\frac{\Delta E_{rad}^{(1)}}{E}\approx \frac{\alpha _{s}C_{F} \mu_D^2
L^2}{4 l_g E}[ln\frac{2E}{\mu_D^2 L}-0.048]\, , \label{derad}
\end{equation}

\begin{equation}
\frac{\Delta E_{abs}^{(1)}}{E}\approx -\frac{\pi\alpha _{s}C_{F}
}{3}\frac{LT^2}{l_g E^2}[ln\frac{\mu_D^2
L}{T}-1+\gamma_E-\frac{6\zeta^{'}(2)}{\pi^2}]\, . \label{deabs}
\end{equation}

As shown in Eq.(\ref{derad}), the analytic approximation of the
energy loss by stimulated emission in a static medium is
proportional to $L^2ln(\frac{1}{L})$. We fit the curve from our
numerical calculation in a chemical non-equilibrated medium, it is
shown as in Fig. \ref{fig:deradcompare} that the energy loss by
stimulated emission is proportional to $L$ approximately by taking
into account the parton evolution of the chemical non-equilibrated
medium and Bjorken expanding medium. In a static medium it is shown
that the energy gain $\Delta E_{abs}^{(1)}$ in Eq. (\ref{deabs}),
from thermal absorption is linear distance dependence if assuming
$\mu_D^2 L/T>>1$, and become a quadratic distance dependence if
assuming $\mu_D^2 L/T<<1$. In our case for a chemical
non-equilibrated medium here, from the curve fitting it is shown as
in Fig. \ref{fig:deabscompare} that the energy gain keep the linear
distance dependence as $\mu_D^2 L/T>>1$, but the distance dependence
becomes $L^{0.2}$ as $\mu_D^2 L/T<<1$ instead of $L^2$-dependence in
static medium case and the proportionality factor is much smaller
than that in the static medium.

As shown in Fig. \ref{fig:deradcompare}, the energy loss without
thermal absorption in a chemical non-equilibrated medium with
initial temperature $T_0=550$ $MeV$, initial fugacity
$\lambda_{g0}=0.3$ is larger than that in a static medium with
$T=300$ $MeV$, $\mu_D=0.5$ $GeV$ and the mean free path of the jet
$l_{g}=1$ $fm$. But it is less than that in the Bjorken expanding
medium with $T=420$ $MeV$. Although the opacity in the Bjorken
expanding medium is less than that in the chemical non-equilibrated
expanding medium, the Debye screening mass is larger, this leads to
larger $q_{\perp max}$ and larger energy loss per collision. We also
calculated the energy loss without thermal absorption in the Bjorken
medium with the same initial temperature $T_0=550$ $MeV$ as in
chemical non-equilibrated medium to study difference between Bjorken
expansion and chemical non-equilibrated expansion medium. It is
shown that the energy loss without thermal absorption in the Bjorken
medium is larger than that in the chemical non-equilibrated medium
due to that temperature decreases more rapidly in the chemical
non-equilibrated medium than the Bjorken expanding medium. As shown
in Fig. \ref{fig:deabscompare}, the energy gain in the chemical
non-equilibrated medium is a bit larger than that in the Bjorken
expanding medium with $T_0=420$ $MeV$ due to the larger temperature
in the chemical non-equilibrated medium.

\begin{figure}
\includegraphics[width=8.6cm]{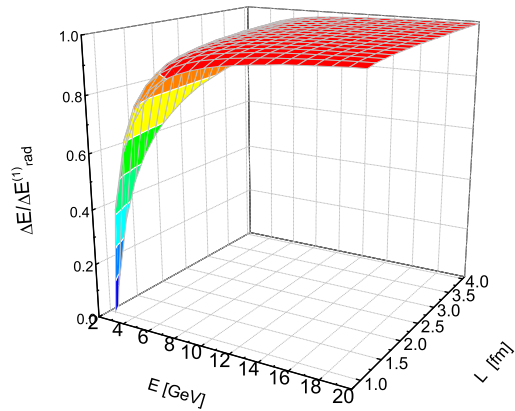}
\caption[a]{The ratio of effective parton energy loss with and
without absorption as a function of parton energy $E$ and
propagating distance $L$.} \label{fig:ratel2}
\end{figure}

\begin{figure}
\includegraphics[width=8.9cm]{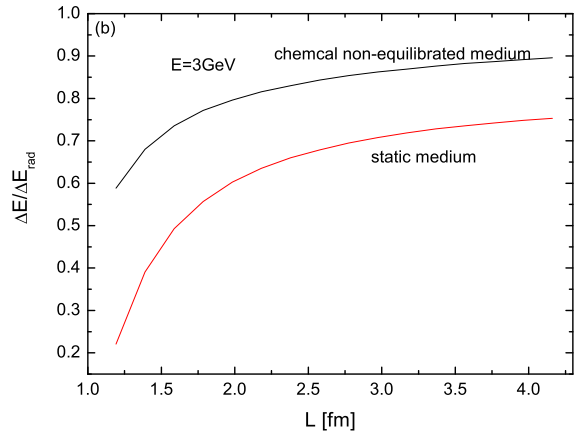}
\caption[a]{The ratio of effective parton energy loss with and
without absorption as a function of propagating distance $L$ at
fixed $E=3$ $GeV$ in a static medium and chemical non-equilibrated
hot medium cases.} \label{fig:rateele3.EPS}
\end{figure}

Fig. \ref{fig:ratel2} shows the ratio between parton energy loss
with thermal absorption $\Delta E=\Delta E_{1rad}+\Delta
E_{1abs}+\Delta E_{0abs}$ and the radiative energy loss $\Delta
E_{1rad}$ without thermal absorption as functions of parton energy
value $E$ and propagating distance $L$ in a chemical
non-equilibrated medium. It is shown that the energy gain via
absorption is important for intermediate parton energy and small
propagating distance. The absorption can be neglected either for
asymptotically large parton energy or large values of $L$. The
comparison between the ratios in a static medium and in a chemical
non-equilibrated hot medium at fixed parton energy $E=3$ $GeV$ is
shown in Fig.\ref{fig:rateele3.EPS}. We can see that the ratio of
the calculated parton energy loss with and without thermal
absorption in a static medium($T=550$ $MeV$, $\mu_D=1$ $GeV$)
considered in Ref. \cite{Wang:2001sf} is less than that in a
chemical non-equilibrated hot medium. This result implies that the
energy gain via absorption is more important in a static medium than
that in a chemical non-equilibrated medium at large distance $L$ and
intermediate jet energy region.

\section{Conclusion}
\label{conc-sec}

In summary, we have determined the initial conditions for central
Au-Au collisions at RHIC. It is shown that the initial temperature
is 550$MeV$, the initial fugacity $\lambda_g$ for gluons is 0.3 and
$\lambda_q$ for quarks is 0.06 for the chemical non-equilibrated
expanding medium. The initial temperature is $420$ $MeV$ for the
Bjorken expanding medium. We obtain that partons cool faster than
predicted by Bjorken with the proper time. This chemical
equilibration in a hot medium lead to that the Debye mass and the
inverse mean free-path decrease with increasing the proper time, and
the opacity for a energetic parton in a chemical non-equilibrated
medium increases with incraesing the propagating distance $L$. The
Debye mass and the mean free-path in the Bjorken expanding medium is
larger than that in the chemical non-equilibrated expanding medium.
We found that the opacity in the chemical non-equilibrated medium is
larger than that in the static medium at temperature $300$ $MeV$ and
that in the Bjorken expanding medium, but is less than that in a
static medium at $T=550$ $MeV$.

The parton evolution affect the energy loss with detailed balance.
The propagating distance $L$ dependence of parton energy loss with
detailed balance is determined. The net effect of stimulated gluon
emission and thermal absorption are considered for an energetic
parton propagating inside the QGP medium. It is shown that, by
taking into account the evolution of the temperature and fugacity,
the energy loss from stimulated emission is proportional to $L$ in
the chemical non-equilibrated medium and Bjorken expanding medium
rather than $L^2$-dependence on the propagating distance in the
static medium. The energy loss in the chemical non-equilibrated
medium is a bit less than that in Bjorken expanding medium. Because
of the higher temperature in the chemical non-equilibrated medium,
the energy gain without rescattering to the first order opacity in
the chemical non-equilibrated medium are larger than those in the
Bjorken medium. The relative reduction of the parton energy loss is
significant at intermediate parton energy E and small propagating
distance $L$ in the chemical non-equilibrium medium. The evolution
of the medium modifies the jet energy loss in the intermediate
energy region and affect the shape of suppression intermediate high
$P_{T}$ hadrons spectrum.

\section*{Acknowledgments}
We acknowledge helpful discussions with X.N. Wang, M. Gyulassy and
C. Greiner. This work was supported by NSFC of China under Projects
No. 10825523, No. 10635020 and No. 10875052, by MOE of China under
Projects No. IRT0624, and by MOE and SAFEA of China under Project
No. PITDU-B08033. L.C. thanks DAAD foundation for their support.

\end{document}